\documentstyle{iopart}

\input epsf.tex

\def\spose#1{\hbox to 0pt{#1\hss}}
\def\lta{\mathrel{\spose{\lower 3pt\hbox{$\mathchar"218$}}
     \raise 2.0pt\hbox{$\mathchar"13C$}}}
\def\gta{\mathrel{\spose{\lower 3pt\hbox{$\mathchar"218$}}
     \raise 2.0pt\hbox{$\mathchar"13E$}}}

\def\eg{{\it e.g.,\,}}
\def\ie{{\it i.e.,\,}}
\def\etal{{\it et al.\,}}

\def\be{\begin{equation}}
\def\ee{\end{equation}}
\def\bea{\begin{eqnarray}}
\def\eea{\end{eqnarray}}
\def\lbl{\label}
\def\hm{{\cal H}^3}

\def\xic{\xi^{c}_\Phi({\bf x}, {\bf x}^\prime)}
\def\xico{\xi^{c}_\Phi({\bf x}, {\bf x})}
\def\xiu{\xi^{u}_\Phi({\bf x}, {\bf x}^\prime)}
\def\xiug{\xi^{u}_\Phi({\bf x}, \gamma {\bf x}^\prime)}

\def\xiur{\xi^{u}_\Phi(r)}
\def\xiurj{\xi^{u}_\Phi(r_j)}

\def\dT{\frac{\Delta T}{T}}

\def\tauls{\tau_{\sc ls}}
\def\chiH{{\chi_{\!{}_{\rm H}}}}

\newcommand{\chix}[1]{{\chi_{\!{}_{\rm #1}}}}

\def\today{\ifcase\month\or
 January\or February\or March\or April\or May\or June\or
 July\or August\or September\or October\or November\or December\fi
 \space\number\day, \number\year}

\begin{document}

\title
{Computing CMB anisotropy in Compact Hyperbolic spaces}
\author{J.Richard Bond, Dmitry Pogosyan and Tarun Souradeep}
\address{Canadian Institute for Theoretical Astrophysics,\\ 
University of Toronto, ON M5S 3H8, Canada}

\begin{abstract}
 The measurements of CMB anisotropy have opened up a window for
probing the global topology of the universe on length scales
comparable to and beyond the Hubble radius. For compact topologies,
the two main effects on the CMB are: (1) the breaking of statistical
isotropy in characteristic patterns determined by the photon geodesic
structure of the manifold and (2) an infrared cutoff in the power
spectrum of perturbations imposed by the finite spatial extent.  We
present a completely general scheme using the {\em regularized method
of images} for calculating CMB anisotropy in models with nontrivial
topology, and apply it to the computationally challenging compact
hyperbolic topologies. This new technique eliminates the need for the
difficult task of spatial eigenmode decomposition on these spaces. We
estimate a Bayesian probability for a selection of models by
confronting the theoretical pixel-pixel temperature correlation
function with the {\sc cobe--dmr} data. Our results demonstrate that
strong constraints on compactness arise: if the universe is small
compared to the `horizon' size, correlations appear in the maps that
are irreconcilable with the observations. If the universe is of
comparable size, the likelihood function is very dependent upon
orientation of the manifold {\it wrt} the sky. While most orientations
may be strongly ruled out, it sometimes happens that for a specific
orientation the predicted correlation patterns are preferred over the
conventional infinite models.
\end{abstract}

\section{Introduction}

The remarkable degree of isotropy of the cosmic microwave background
(CMB) points to homogeneous and isotropic Freidmann-Robertson-Walker
(FRW) models for the universe. The underlying Einstein's equations of
gravitation are purely local, completely unaffected by the global
topological structure of space-time.  In fact, in the absence of
spatially inhomogeneous perturbations, a FRW model predicts an
isotropic CMB regardless of the global topology.

The observed large scale structure in the universe implies spatially
inhomogeneous primordial perturbations exist which gave rise to the
observed anisotropy of the CMB. The global topology of the universe
does affect the local observable properties of the CMB anisotropy. In
compact universe models, the finite spatial size usually precludes the
existence of primordial fluctuations with wavelengths above a
characteristic scale related to the size of the universe.  As a
result, the power in the CMB anisotropy is suppressed on large angular
scales. Another consequence is the breaking of statistical isotropy in
characteristic patterns determined by the photon geodesic structure of
the manifold. One can search for such patterns statistically in the
COBE maps, and to the extent that they are not there, one can
constrain the size of the universe and its topology.

Much recent astrophysical data suggest the cosmological matter density
parameter, $\Omega_0$, is subcritical~\cite{opencase}. If
a (possibly varying) cosmological constant is absent or insufficient to
bring the total density to the critical value, this would
imply a hyperbolic spatial geometry for the universe (commonly
referred to as the `open' universe in the cosmological
literature). The topologically trivial (simply connected) hyperbolic
3-space, $\hm$, is non-compact and has infinite size.  There are
numerous theoretical motivations, however, to favor a spatially
compact universe~\cite{motive}.  To reconcile a compact universe with
a flat or hyperbolic geometry, consideration of spaces with non
trivial topology (non simply connected spaces) is required. A compact
cosmological model can be constructed by identifying points on the
standard infinite flat or hyperbolic FRW spaces by the action of a
suitable discrete subgroup, $\Gamma$, of the full isometry group, $G$,
of the FRW space. The infinite FRW spatial hypersurface is the {\em
universal cover}, tiled by copies of the compact space (most
appropriately represented as the {\em Dirichlet domain} with the
observer at its {\em basepoint}).  Any point ${{\bf x}}$ of the
compact space has an image ${{\bf x}}_i = \gamma_i {{\bf x}}$ in each copy
of the Dirichlet domain on the universal cover, where $\gamma_i \in
\Gamma$.

The hyperbolic manifold, $\hm$, can be viewed as a hyperbolic section
embedded in four dimensional flat Lorentzian space. The isometry group
of $\hm$ is the group of rotations in the four space -- the proper
Lorentz group, $SO(3,1)$. A compact hyperbolic (CH) manifold is then
completely described by a discrete subgroup, $\Gamma$, of the proper
Lorentz group, $SO(3,1)$. The Geometry Centre at the University of
Minnesota has a large census of CH manifolds and public domain
software SnapPea~\cite{Minn}. We have adapted this software to tile
$\hm$ under a given topology using a set of generators of $\Gamma$.
The tiling routine uses the generator product method and ensures that
all distinct tiles within a specified tiling radius are obtained.

A CH manifold, ${\cal M}$, is characterized by a dimensionless number,
${\cal V}_{\!\cal M}\equiv V_{\!\cal M}/d_c^3$, where $V_{\!\cal M}$
is the volume of the space and $d_c$ is the curvature radius
\cite{Thur7984}. There are a countably infinite number of CH manifolds
with no upper bound on ${\cal V}_{\!\cal M}$.  The smallest CH
manifold discovered so far has ${\cal V}_{\!\cal M}
=0.94$~\cite{smallestCH}. ~\footnote{ The volume of CH manifolds is
bounded from below and the present theoretical lower bound stands at
${\cal V}_{\!\cal M} \ge 0.167$~\cite{gab_mey96}. There exist sharper
lower bounds within subclasses of CH manifolds under restrictions on
topological invariants~\cite{minvol}. It has been conjectured that the
smallest known manifold is in fact the smallest
possible~\cite{smallestCH}.} The Minnesota census lists several
thousands of these manifolds with ${\cal V}_{\!\cal M}$ up to $\sim
7$. In the cosmological context, the physical size of the curvature
radius $d_c$ is determined by the density parameter and the Hubble
constant $H_0$: $d_c=(c/H_0)/\sqrt{1-\Omega_0}$. The physical volume
of the CH manifold with a given topology, \ie a fixed value of
$V_{\!\cal M}/d_c^3$, is smaller for smaller values of $\Omega_0$. Two
quantities which characterize linear dimensions of the Dirichlet domain
are $R_>$ and $R_<$, the radii of circumscribing and inscribing
spheres, respectively.

In the standard picture, the CMB that we observe is a Planckian
distribution of relic photons which decoupled from matter at a
redshift $\approx 1100$. These photons have freely propagated over a
distance $ R_{\sc ls} \approx 2 d_c\, {\rm arctanh} \sqrt{1-\Omega_0}
$, comparable to the ``horizon'' size.  For the adiabatic fluctuations
we consider here, the dominant contribution to the anisotropy in the
CMB temperature measured with wide-angle beams ($\theta_{\sc fwhm}
\gta 2^\circ \Omega_0^{1/2}$) comes from the cosmological metric
perturbations through the Sachs-Wolfe effect.

The adiabatic cosmological metric perturbations can be expressed in
terms of a scalar gravitational potential, $\Phi({\bf x},\tau)$.  The
dynamical equation for the gravitational potential allows for
separation of the spatial and temporal dependence,\footnote{At the
scales appropriate to CMB anisotropies, damping effects on $\Phi$ can be
neglected.}  $\Phi({\bf x},\tau) = F(\tau) \Phi({\bf x},\tauls)$,
where $F(\tau)$ encodes the time dependence of the metric
perturbations and $\Phi({\bf x},\tauls)$ is the field configuration on
the three-hypersurface of constant time $\tau=\tauls$ when the last
scattering of CMB photons took place. We shall study open,
$\Omega_0<1$, models with zero cosmological constant where in the
matter dominated phase,~\cite{mukh92}

\begin{equation}
F(\tau) = \frac{5(\sinh^2\tau -3 \tau\sinh\tau +4\cosh\tau -4)}
{(\cosh\tau -1)^3}\,.
\lbl{Feta}
\end{equation}
Here and further on we use dimensionless conformal time $\tau$
expressed in units of the curvature radius. A non-zero cosmological
constant can be trivially incorporated in our analysis by using the
appropriate solution for $F(\tau)$.

We write the Sachs-Wolfe formula for the CMB temperature fluctuation,
$\Delta T(\hat q)$, in a direction $\hat q$, in the form
\begin{equation}
\fl \dT(\hat q) = \frac{1}{3} \Phi(\hat q\chiH,\tauls) +
2 \int_{0}^\chiH d\chi f(\chi)
 \Phi(\hat q\chi,\tauls)\,, \quad 
 f(\chi)=\frac{d}{d\tau} F(\tau)\bigg|_{\tau=\chiH-\chi}\,,
\lbl{dTSW}
\end{equation}
where $\chi$ is the affine parameter along the photon path from
$\chi=0$ at the observer position to $\chiH=R_{\sc ls}/d_c$. The first term
is called the {\em surface} or ``naive'' Sachs-Wolfe effect (NSW). The
second term, which is nonzero only if $\Phi$ varies with time between
$\tauls$ and now, is the {\em integrated} Sachs-Wolfe effect
(ISW). The angular correlation between the CMB temperature
fluctuations in two directions in the sky is then given by
\begin{eqnarray}
\fl C(\hat q,\hat q^\prime)\equiv \left\langle\dT(\hat q)\dT(\hat
q^\prime)\right\rangle = \frac{1}{9} \langle\Phi(\hat
q\chiH,\tauls)\Phi(\hat q^\prime\chiH,\tauls)\rangle \nonumber\\ \lo
+\frac{2}{3}\int_{0}^\chiH d\chi~f(\chi)~\left[ \langle\Phi(\hat
q\chi,\tauls)\Phi(\hat q^\prime\chiH,\tauls)\rangle +\langle\Phi(\hat
q^\prime\chi,\tauls)\Phi(\hat q\chiH,\tauls)\rangle\right] \nonumber\\
\lo +4\int_{0}^\chiH d\chix{1} ~f(\chix{1})~ \int_{0}^\chiH
d\chix{2}~f(\chix{2})~ \langle\Phi(\hat q\chix{1},\tauls)\Phi(\hat
q^\prime\chix{2},\tauls)\rangle\,.  \lbl{cthetaSW}
\end{eqnarray}
The main point to be noted is that $C(\hat q,\hat q^\prime)$ depends
on the spatial two point correlation function, $\xi_\Phi \equiv
\langle\Phi({\bf x},\tauls)\Phi({\bf x^\prime},\tauls)\rangle $ of
$\Phi$ on the three-hypersurface of last scattering. This is due to
the fact that the equation of motion for $\Phi$ allows a separation of
spatial and temporal dependence.

Although in this work we restrict our attention to the Sachs-Wolfe
effect which dominates when the beam size is large, we should point
out that other effects which contribute to the CMB anisotropy at finer
resolution can also be approximated in terms of spatial correlation of
quantities defined on the hypersurface of last scattering~\cite{us_inprep}.

\section{Method of Images}
\lbl{sec_moi}

As described in the previous section, the angular correlation function
for the CMB temperature anisotropy can be expressed in terms of the
two point spatial correlation function $\xi_\Phi({\bf x}, {\bf
x}^\prime)$ for the gravitational potential field, $\Phi({\bf
x},\tauls )$ defined on the three-hypersurface of last
scattering. The correlation function $\xi_\Phi$ can be expressed
formally in terms of the eigenfunctions, $\Psi_i$ of the Laplace
operator, $\nabla^2$, on the hypersurface (with positive eigenvalues
$k_i^2 \ge 0$), as~\cite{mukh92,bondLH}
\begin{eqnarray} 
\xi_\Phi({\bf x}, {\bf x}^\prime) = \sum_i \Psi_i({\bf
x})\Psi_i^*({\bf x}^\prime) P_\Phi(k_i)\, , \ {\rm where}\ (\nabla^2 +
k_i^2)\Psi_i = 0\,. \lbl{xi1}
\end{eqnarray}
The function $P_\Phi(k_i)$ is the initial power spectrum of the
gravitational potential.

In order to calculate the $\xi_\Phi$ on CH manifolds we have developed
the {\em method of images}~\cite{us_texas} which evades the difficult
problem\footnote{Obtaining closed form expressions for eigenfunctions
may not be possible beyond the simplest topologies. Examples where
explicit eigenfunctions have been used include flat
models~\cite{star_angel,tor_refs,us_torus,janna} and noncompact
hyperbolic space with horn topology~\cite{sok_star75,janna}. The
generically chaotic nature of the geodesic flow on CH manifolds
presents an inherently difficult problem even in terms of a numerical
estimation of eigenfunctions.} of solving for eigenfunctions of the
Laplacian on these manifolds and, in general, on any nonsimply
connected manifold, ${\cal M}$.  The manifold ${\cal M}$ can be
regarded as a quotient of a simply connected manifold ${\cal M}^u$ by
a fixed point free discontinuous subgroup $\Gamma$ of the isometry
group of ${\cal M}^u$, i.e., ${\cal M}= {\cal M}^u/\Gamma$. The
manifold ${\cal M}^u$ is referred to as the {\em universal cover} of
${\cal M}$. The elements $\gamma\in\Gamma$ produce images of the
manifold ${\cal M}$ which tessellate ${\cal M}^u$.

The spectrum of the Laplacian on a compact space (thus with closed
boundary conditions) is a discrete ordered set of eigenvalues $\{
k^2_i\}$ ($k_0^2 = 0$ and $k_i^2 \le k^2_{i+1}$, with multiply repeated
eigenvalues counted separately). The set $\{\Psi_i\}$ is the
corresponding complete orthonormal set of eigenfunctions on ${\cal
M}$.  The eigenfunctions $\Psi_i$ on ${\cal M}$ are also
eigenfunctions on the universal cover, ${\cal M}^u$.  As a result
these eigenfunctions simultaneously satisfy the integral equations for
the correlation functions in both ${\cal M}^u$ and ${\cal M}$,
\begin{eqnarray}
\int_{{\cal M}^u} d{\bf x^\prime}~\xiu~\Psi_i({\bf x}^\prime) =
P_\Phi(k_i)\Psi_i({\bf x})\,,\\ \int_{\cal M} d{\bf
x^\prime}~\xic~\Psi_i({\bf x^\prime}) = P_\Phi(k_i)\Psi_i({\bf x})\,.
\lbl{eigeqn3}
\end{eqnarray}
Using the fact that the $\Psi_i$ are automorphic with respect to $\Gamma$
($\Psi(\gamma{\bf x})= \Psi({\bf x}), \forall \gamma\in\Gamma$), and
that ${\cal M}$ tessellates ${\cal M}^u$,
\begin{eqnarray}
\int_{{\cal M}^u} d{\bf x^\prime}~\xiu \Psi_i({\bf x^\prime}) &=
\sum_{\gamma\in\Gamma}\int_{\cal M} d{\bf x^\prime}~ \xiug~ \Psi_i({\bf
x^\prime}) \nonumber\\ &= \int_{\cal M} d{\bf
x^\prime}~\left[~\sum_{\gamma\in\Gamma}^\prime \xiug ~\right]
\Psi_i({\bf x^\prime}), \lbl{kernelsum}
\end{eqnarray}
where $\sum^\prime $ denotes a possible need for regularization at the
last step when the order of integration and summation is reversed.
The regularization is achieved through a counterterm which removes the
contribution of the homogeneous $k^2=0$ mode.  Using
eqs.~(\ref{eigeqn3}) and (\ref{kernelsum}) leads to the main equation
of our {\em method of images}~\cite{us_inprep} expressing the
correlation function on a compact space (and more generally, any
non-simply connected space) as a sum over the correlation function on
its universal cover calculated between ${\bf x}$ and the images
$\gamma{\bf x^\prime}$ ($\gamma\in\Gamma$) of ${\bf x^\prime}$:
\begin{eqnarray}
\xic &= \sum_{\gamma\in\Gamma}^\prime \xiug \nonumber \\ &=
\sum_{\gamma\in\Gamma} \xiug -\frac{1}{V_{\!\cal M}}\int_{{\cal
M}^u}d{\bf x^\prime}~\xiu \,.  \lbl{moi1}
\end{eqnarray}

\section{Compact Hyperbolic Models}

For cosmological CH models, ${\cal M}^u \equiv \hm$, the three
dimensional hyperbolic (uniform negative curvature) manifold. The
local isotropy and homogeneity of $\hm$ implies $\xiu$ depends only on
the proper distance, $r\equiv d({\bf x},{\bf x^\prime})$, between the
points ${\bf x}$ and ${\bf x^\prime}$.  The eigenfunctions on the
universal cover are of course well known for all homogeneous and
isotropic models~\cite{har67}. Consequently $\xiu$ can be obtained
through equation (\ref{xi1}).  The initial power spectrum $P_\Phi(k)$
is believed to be dictated by an early universe scenario for the
generation of primordial perturbations. We assume that the initial
perturbations are generated by quantum vacuum fluctuations during
inflation. This leads to
\begin{equation} 
\xiu \equiv \xiur = \int_0^\infty
\frac{d\beta~\beta}{(\beta^2 +1)} ~\frac{\sin(\beta r)}{\beta \sinh
r}~~{\cal P}_\Phi(\beta) \, , \lbl{xiur}
\end{equation}
where $\beta \equiv \sqrt{(k d_c)^2 -1}$ and ${\cal
P}_\Phi(\beta)\equiv \beta(\beta^2+1) P_\Phi(k)/(2\pi^2)$ is the mean
square fluctuation in $\Phi$ per logarithmic interval of $k$. We
consider the simplest inflationary models, where ${\cal P}_\Phi(k)$ is
approximately constant in the subcurvature sector ( $k d_c > 1$). This
is the generalization of the Harrison-Zeldovich spectrum in spatially
flat models to hyperbolic spaces~\cite{lyt_stew90,rat_peeb94}.
Sub-horizon vacuum fluctuations during inflation are not expected to
generate supercurvature modes, those with $k d_c <1$, which is why
they are not included in eq.(\ref{xiur}). Indeed, since $H^2 > 1/(a
d_c)^2$, for modes with $k d_c < 1$ we always have $k/(aH) < 1$ so
inflation by itself does not provide a causal mechanism for their
excitation. Moreover, the lowest non-zero eigenvalue, $k_1>0$ in
compact spaces provides an infra-red cutoff in the spectrum which can
be large enough in many CH spaces to exclude the supercurvature sector
entirely ($k_1 d_c > 1$).  (See $\S$~\ref{ssec_res_powspec} and
\cite{us_inprep}.)  Even if the space does support supercurvature
modes, some physical mechanism needs to be invoked to excite them, \eg
as a byproduct of the creation of the compact space itself, but which
could be accompanied by complex nonperturbative structure as well.  To
have quantitative predictions for $P_\Phi(k)$ would require addressing
this possibility in a full quantum cosmological context.  We note that
our main conclusions regarding peculiar correlation features in the
CMB anisotropy (see $\S$~\ref{ssec_res_cmbcorr}) would qualitatively
hold even in the presence of supercurvature modes.

\subsection{Numerical implementation of method of images}

Although equation (\ref{moi1}) encodes the basic formula for
calculating the correlation function, it is not numerically
implementable as is. Both the sum and the integral in equation
(\ref{moi1}) are divergent and the difference needs to be taken as a
limiting process of summation of images and integration up to a finite
distance $r_*$:
\begin{equation}
\fl\xic = \lim_{r_*\to\infty} \left[ \sum_{r_j < r_*} \xiurj - 
\frac{4\pi}{V_{\!\cal M}} \int_0^{r_*} dr ~\sinh^2r ~\xiur\right],~~~~
r_j = d({\bf x}, \gamma_j {\bf x}^\prime)\, . 
\lbl{moi_final}
\end{equation}
The volume element in the integral corresponds to $\hm$.  In
\cite{us_inprep} we conduct an instructive analytical study of the
regularization procedure by varying an artificial infrared cutoff.

In hyperbolic spaces, the number of images within a radius $r_*$ grows
exponentially with $r_*$ and it is not numerically feasible to extend
direct summation to large values of $r_*$. The presence of the
counterterm, however, besides regularizing, significantly improves
convergence. This can be intuitively understood as follows~: $\xiurj$
represents a sampling of a smooth function at discrete points
$r_j$. In a distant radial interval $[r,r+dr],~r \gg R_>, ~dr \sim
R_>$ there are approximately $ (4\pi/V_{\!\cal M})~\sinh^2r~dr$
images.  The sum, $\sum_{r_j} \xiurj$, within this interval is similar
to the (Monte-Carlo type) estimation of the integral, therefore one
may approximate the sum over all distant images beyond a radius $r_*$
by an integral to obtain
\begin{equation}
\tilde\xic =  \sum_{r_j < r_*} \xiurj + 
\frac{4\pi}{V_{\!\cal M}} \int_{r_*}^\infty dr~ \sinh^2r ~\xiur\,.
\lbl{moi3}
\end{equation}
The tilde on $\xic$ denotes the fact that it is approximate and
unregularized. Subtracting the integral $(4\pi/V_{\!\cal
M})\int_0^\infty dr ~\sinh^2r ~\xiur $ as dictated by the
regularization eq.(\ref{moi1}), we recover the finite $r_*$ term of
the limiting sequence in eq.(\ref{moi_final}).  This demonstrates that
even at a finite $r_*$, in addition to the explicit sum over images
with $r_j<r_*$, the expression for $\xi^c_\Phi$ in
eq.(\ref{moi_final}) contains the gross contribution from all distant
images, $r_j > r_*$.  Numerically we have found it suffices to
evaluate the right hand side in eq.(\ref{moi_final}) up to $r_*$ about
$4$ to $5$ times the domain size $R_>$ to obtain a convergent result
for $\xic$.

 After this project was completed, we learned that estimation
(\ref{moi3}) is similar to the remainder term obtained in
~\cite{aur_stein92} to incorporate the gross contributions of long
periodic orbits in evaluating the ``Selberg trace formula''(\eg
\cite{chav84}) in the quantum particle problem on 2D CH surfaces.
Whereas the focus in \cite{aur_stein92} was an accurate computation of
eigenvalues for determining the quantum energy levels of the
classically chaotic system, our focus is on correlation function
evaluation in 3D. We emphasize that it is the regularized procedure
(\ref{moi_final}) following directly from eq.(\ref{moi1}) which is
central to the success of our method. The form of the unregularized
correlation function, $\tilde\xic$, in eq.(\ref{moi3}) suggested by
the approximate treatment of distant images discussed above, in fact,
also follows from the form of the regularizing counterterm in
eq.(\ref{moi1}).

\begin{figure}[htb]
\centerline{\epsfxsize=3.75in\epsfbox{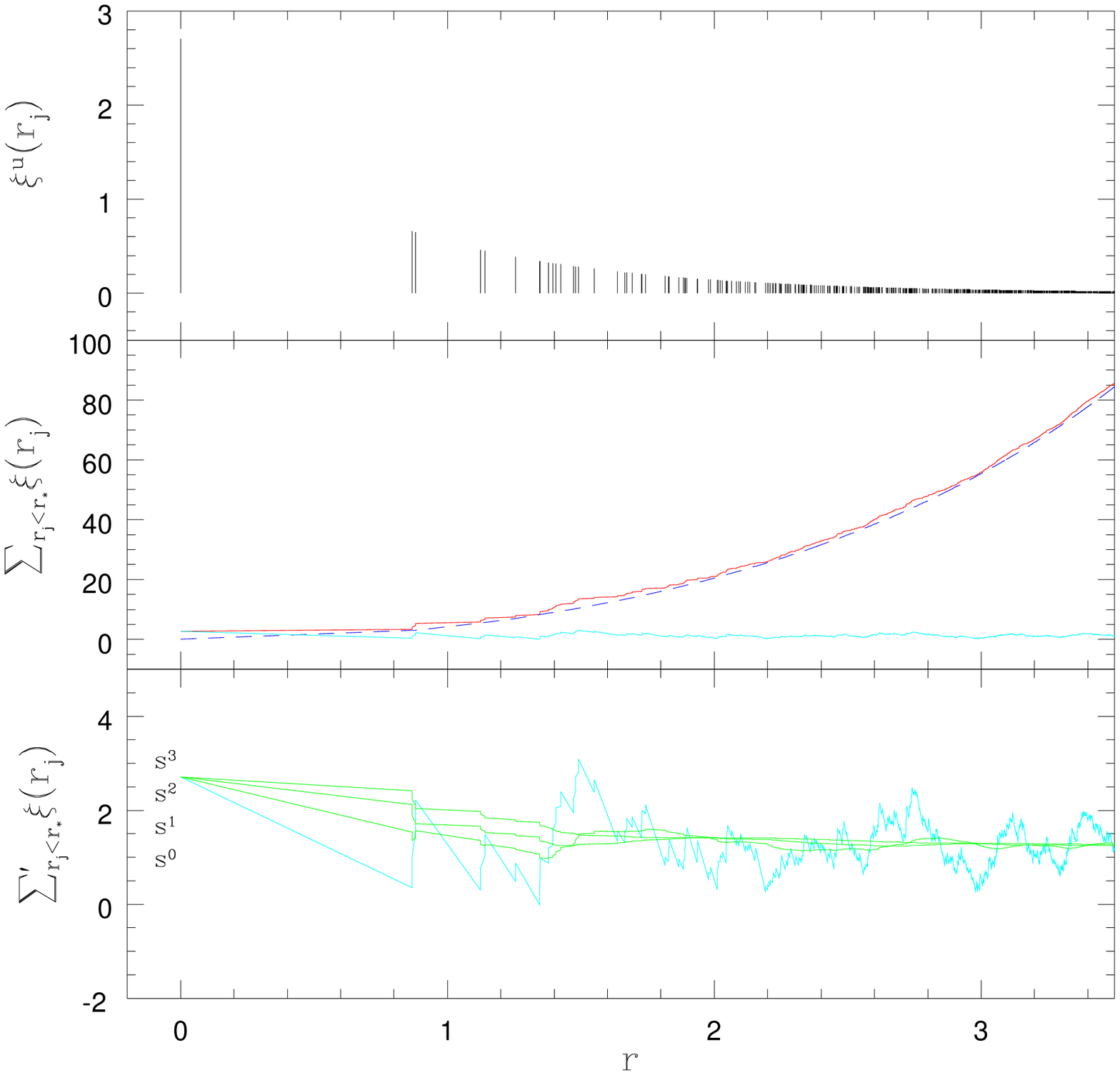}}
\caption{This illustrates the regularization of the correlation
function $\xic$. The example shown is the correlation function at
zero-point separation at some point on the CH manifold m004(-5,1).
The topmost panel shows the contributions to $\xic$ from close images
and the distances to the images.  The upper solid curve in the middle
panel shows the cumulative build up of $\sum_{r_j < r_*} \xiurj$ with
successive addition of distant images. The dashed curve is the
regularizing counterterm required to remove the zero-mode contribution
and the lower solid curve shows the cumulative value of regularized
$\xic$ which fluctuates around the true value after sufficient number
of images have been added. The bottom panel demonstrates that this
residual jitter ($s^0$) in the estimate of $\xic$ can be removed by
Cesaro resummation. The sequence of lines, $s^1,s^2$\& $s^3$ shows the
result of applying first, second and third order Cesaro
resummation. The accuracy at the second order is usually sufficient. }
\label{fig_moi} 
\end{figure}

\Fref{fig_moi} illustrates the steps involved in implementing the
regularized method of images.  The value of $\xi^c_\Phi$ as a function
of $r_*$ has some residual jitter, which arises because of the boundary
effects due to the sharp ``top-hat'' averaging over a spherical ball 
chosen for the counterterm in eq.(\ref{moi_final}).  This can be
smoothed out by resummation techniques~\cite{hardy}.  We use Cesaro
resummation for this purpose.

\subsection{Computing the CMB anisotropy}

For Gaussian perturbations, the angular correlation function, $C(\hat
q,\hat q^\prime)$, completely encodes the CMB anisotropy predictions
of a model. To confront observations, the celestial sphere is
discretized into $N_p$ pixels labeled by $p$, with $N_p$ determined by
the angular resolution of the measurements. We test the models against
the four-year {\sc cobe--dmr} data. The six {\sc cobe--dmr} maps
\cite{dmr4} are first compressed into a (A+B)(31+53+90 GHz)
weighted-sum map, with the customized Galactic cut advocated by the
DMR team, basically at $\pm 20^\circ$ but with extra pixels removed in
which contaminating Galactic emission is known to be high, and with
the dipole and monopole removed. Although one can do analysis with the
map's $(2.6^\circ)^2$ pixels, this ``resolution 6'' pixelization of
the quadrilateralized sphere is oversampled relative to the COBE beam
size, and there is no effective loss of information if we do further
data compression by using ``resolution 5'' pixels, $(5.2^\circ)^2$
\cite{bdmr294}.  The celestial sphere is then represented by
$N_p=1536$ pixels before the Galactic cut, with $N_p=999$ pixels
remaining after the cut is made.

Full Bayesian linear comparison of the model with the data requires
computation of the $N_p\times N_p$ pixel-pixel correlation matrix,
$C_{Tpp^\prime}\equiv C(\hat q_p,\hat q_{p^\prime})$.  The expression
for $C(\hat q,\hat q^\prime)$ in eq.~(\ref{cthetaSW}) involves an
integral along the line of sight from the observer to the surface of
last scattering -- the integrated Sachs--Wolfe term (ISW). We find
that a simple integration rule using $N_L$ points along each line of
sight gives accurate results when the points are spaced at equal
increments of $F(\tau)$, which is determined by
$\Omega_0$. Consequently, to get the full $C_{Tpp^\prime}$ matrix we
need to evaluate the correlation function, $\xic$, between $N_p(N_p+1)
N_L^2/2$ pairs of points on the constant-time hypersurface of last
scattering.

To calculate $C_{Tpp^\prime}$ in the standard infinite open
cosmological models with this real-space integration to an accuracy
comparable to that of the traditional evaluation in \mbox{${\bf
k}$--space}, $N_L \sim 10$ is sufficient. Moreover, the real space
integration is much faster in terms of CPU time. For the method to
remain accurate in compact models, $N_L$ should, of course, exceed the
number of the times a typical photon path crosses the compact
Universe.  We found $N_L \sim 10$ is still enough for the models we
have analyzed so far.

\section{Results} 
\lbl{sec_res}

In $\S$~\ref{ssec_res_powspec}, we discuss the power spectrum of
fluctuations in CH spaces.  In $\S$~\ref{ssec_res_cmbcorr}, we discuss
how the tessellation of $\hm$ by the finite domains is reflected in
the CMB anisotropy correlation function.  In $\S$~\ref{ssec_res_prob},
we present the results of full Bayesian probability analyses of large
angle CMB anisotropy predictions for a selected set of CH models.

\subsection{Power spectrum }
\lbl{ssec_res_powspec}

When going to a compact space from its universal cover, the power
which is spread out in a continuum in $k$-space gets redistributed
and accumulates in peaks at the discrete eigenvalues allowed in the
compact space. For example, in the well known case of $T^3$ (flat
3-torus), the power spectrum is a series of delta functions at $k =
2\pi \sqrt{(n_x/L_x)^2+(n_y/L_y)^2+(n_z/L_z)^2}$, where $n_x$, $n_y$
and $n_z$ are integers.

By contrast, for CH manifolds precise spectra of eigenvalues of the
Laplacian are not known.  We can obtain some information about the
spectrum by applying the method of images to each $\beta$ mode in the
integrand ({\ref{xiur}) for the correlation function at zero-lag
$\xico$. The quantity ${\cal P}_\Phi^c (k,{\bf x}) $ estimated in this way
is defined by $\xico=\int_0^{\infty} d \ln (k) {\cal P}_\Phi^c (k,{\bf x})
$ and describes the contribution to $\xico$ from a unit logarithmic
band $d \ln (k)=1$.  For a discrete spectrum, ${\cal P}_\Phi^c (k,{\bf
x})=\sum_i \delta(k-k_i) {\cal P}_\Phi(k) \sum_j {\left|\Psi_j({k,\bf
x})\right|}^2 $ (here, differently from eq.(\ref{xi1}), $i$ enumerates
only the distinct eigenvalues and $j$ labels the degenerate
eigenfunctions belonging to the same eigenvalue; $\delta$ is the
radial delta-function). Indeed, in the formalism of images the
singular delta-functions in the spectrum are recovered, in principle,
by the precise cancellation of the contributions from {\em all}
images. Determination of the spectrum ${\cal P}_\Phi^c (k) $, in this
sense, is a far more difficult task than that of the correlation
function $\xi^c_{\Phi}$, which is non-singular. Our approximation
technique includes gross integral estimation of the impact of distant
images and results in a spread-out convolved spectral power
distribution. Increasing $r_*$ progressively sharpens spectral
profiles near true positions of discrete eigenvalues~\cite{bal_vor86}.

In general, CH spaces are not only globally anisotropic but also
globally inhomogeneous, even though the universal covering space
(here, ${\cal H}^3$) is homogeneous and isotropic. For ${\cal
P}_\Phi^c (k,{\bf x}) $ in compact space the inhomogeneity manifests
itself in the dependence on ${\bf x}$.  \Fref{fig:kcutoffs} shows a
sample of power spectra for some of the CH manifolds at two random
positions ${\bf x}$.  There is a definite signature of a strong
suppression of power at small $\beta$ in all the cases that we have
explored. Since this suppression is observed at all ${\bf x}$, we
conclude that the small $\beta$ part of the spectrum is absent in the
models considered.  This is qualitatively similar to the infrared
cutoff known for the compact manifolds with flat and spherical
topology.  Quantitatively too, the break appears around $k \sim {\cal
O}(R_>^{-1})$, consistent with intuitive expectations.
 
An infra-red cutoff at the lowest non-zero eigenvalue, $k_1 >0$,
exists for all compact spaces.  The {\em Cheeger's
inequality}~\cite{cheeg70} provides a lower bound on $k_1^2$ for a
compact Riemannian manifold, $M$:
\begin{equation}
k_1^2 \ge \frac{h_C^2}{4}, ~~~~h_C= \inf_{S}~\frac{A(S)}{min
\{V(M_1),V(M_2)\}} \, , 
\lbl{cheeg_ineq}
\end{equation}
where the infimum is taken over all possible surfaces, $S$, that
partition the space, $M$, into two subspaces, $M_1$ and $M_2$, i.e.,
$M=M_1\cup M_2$ and $S=\partial M_1 =\partial M_2$ ($S$ is the
boundary of $M_1$ and $M_2$).  The {\em isoperimetric constant} $h_C$
depends more on the geometry than the topology of the space, with small
values of $h_C$ achieved for spaces having a ``dumbbell-like'' feature
-- a thin bottleneck which allows a partition of the space into two
large volumes by a small-area surface~\cite{cheeg70,bus80}. More
regular compact spaces do not allow eigenvalues which are too small. For
example, the Cheeger limit for all flat $T^3$ manifolds is $k_1 \ge
2/L$, where $L$ is the longest side of the torus.  Although estimation
of $h_C$ is not simple, for any compact space, ${\cal M}$, with
curvature bounded from below there exists a lower bound on $h_C$ in
terms of the diameter of ${\cal M}$, $d_{\!\cal M} \equiv
\sup_{x,y\in{\cal M}}d(x,y)$~\cite{ber80,chav84}; for a 3-dimensional
CH space,
\begin{equation}
k_1 \ge h_C/2 \ge \frac{1}{d_{\!\cal M}}~ {\left[2\int_0^{1/2}\!\!dt
\cosh^2(t)\right]}^{-1}= 0.92/d_{\!\cal M}.  \lbl{cheegmin}
\end{equation}
This result prohibits supercurvature modes for all CH spaces with
$d_{\!\cal M} < 0.92\,d_c$.

There are also upper bounds on $k_1$. The bound, $k_1^2 \le 4 h_C/d_c
+ 10 h_C^2$,~\cite{bus82} does not allow for a firm conclusion that
the space supports supercurvature modes, $k_1 d_c < 1$, unless (using
eq.~(\ref{cheegmin})) $d_{\!\cal M} \ge 10.6\,d_c$ ($R_> \ge
5.3\,d_c$).  Upper bounds based on comparison with the first Dirichlet
eigenvalue on a subdomain of ${\cal M}$ cannot impose $k_1 d_c <1$,
since Dirichlet eigenvalues cannot be less than $d_c^{-1}$.

\begin{figure}[htb]
\centerline{\epsfxsize=4.in\epsfbox{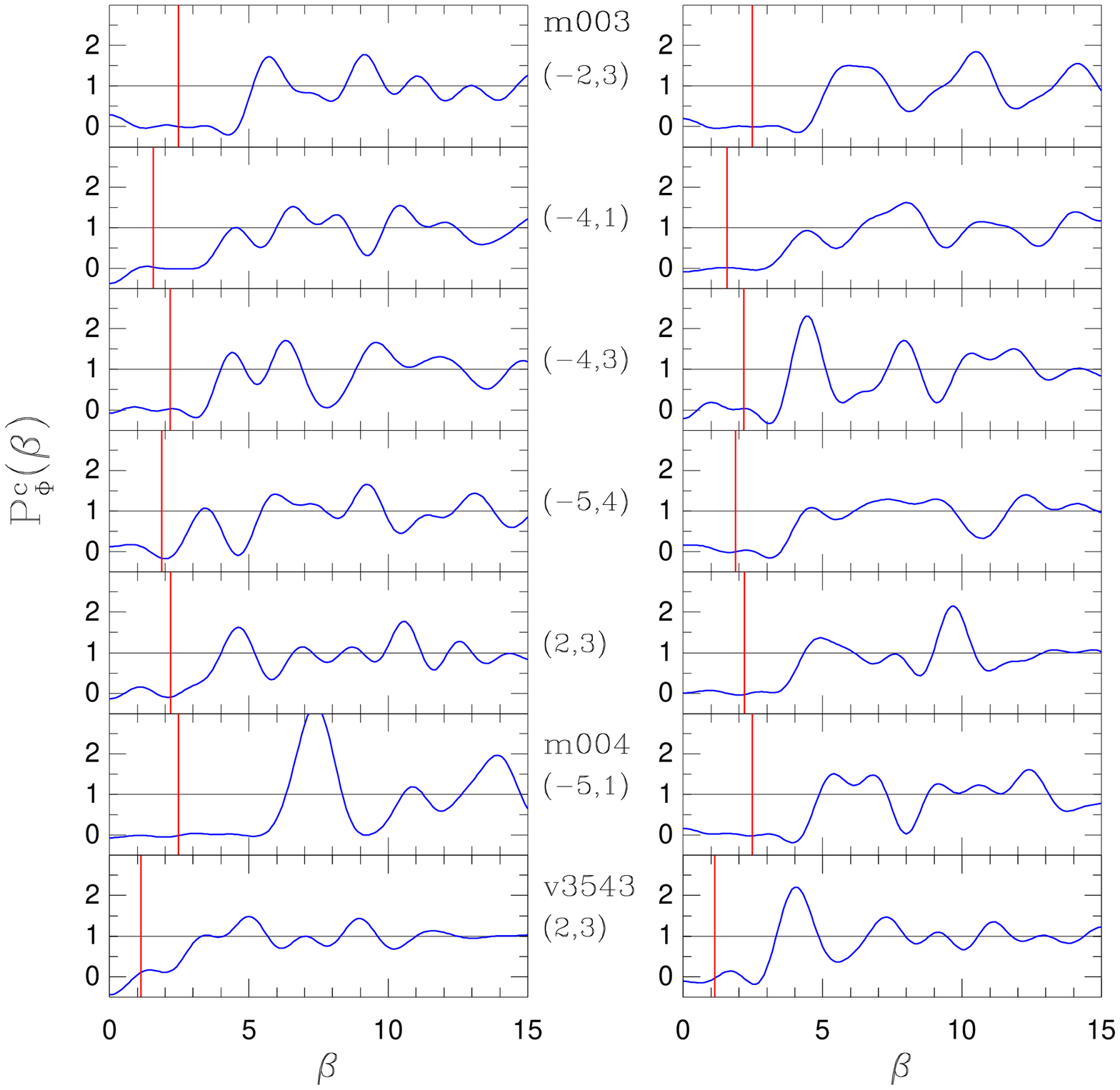}}
\caption{ The primordial power spectrum ${P}^c_{\Phi}( \beta )$
exhibits suppression of long-wave power if the Universe has compact
topology.  ${P}^c_\Phi( \beta )=1$ corresponds to an infinite open
Universe.  Examples of the spectrum cutoff in several CH Universes are
shown.  The first 6 have about the same value of $R_{>}/d_c$, where
$R_{>}$ is the minimal radius of the sphere which completely
encompasses the Dirichlet domain. These spaces have volumes $V/d_c^3$
around unity. The sixth space is m004(-5,1), and the seventh one is
the v3543(2,3) space with volume $V_{\!\cal M}/d_c^3= 6.45$, which are
used as examples of ``small'' and ``large'' spaces in this paper.  The
plots show the integrand for the correlation function $\xic$ at zero
separation, ${\bf x}={\bf x^\prime}$, at two different locations (left
and right panels). This contrasts with isotropic spaces for which the
integrands will be the same. The vertical line at $k=2/R_>$ marks the
value below which there is strong suppression of large-scale power .}
\label{fig:kcutoffs}
\end{figure}

\begin{figure}[htb]
\centerline{\epsfxsize=4.in\epsfbox{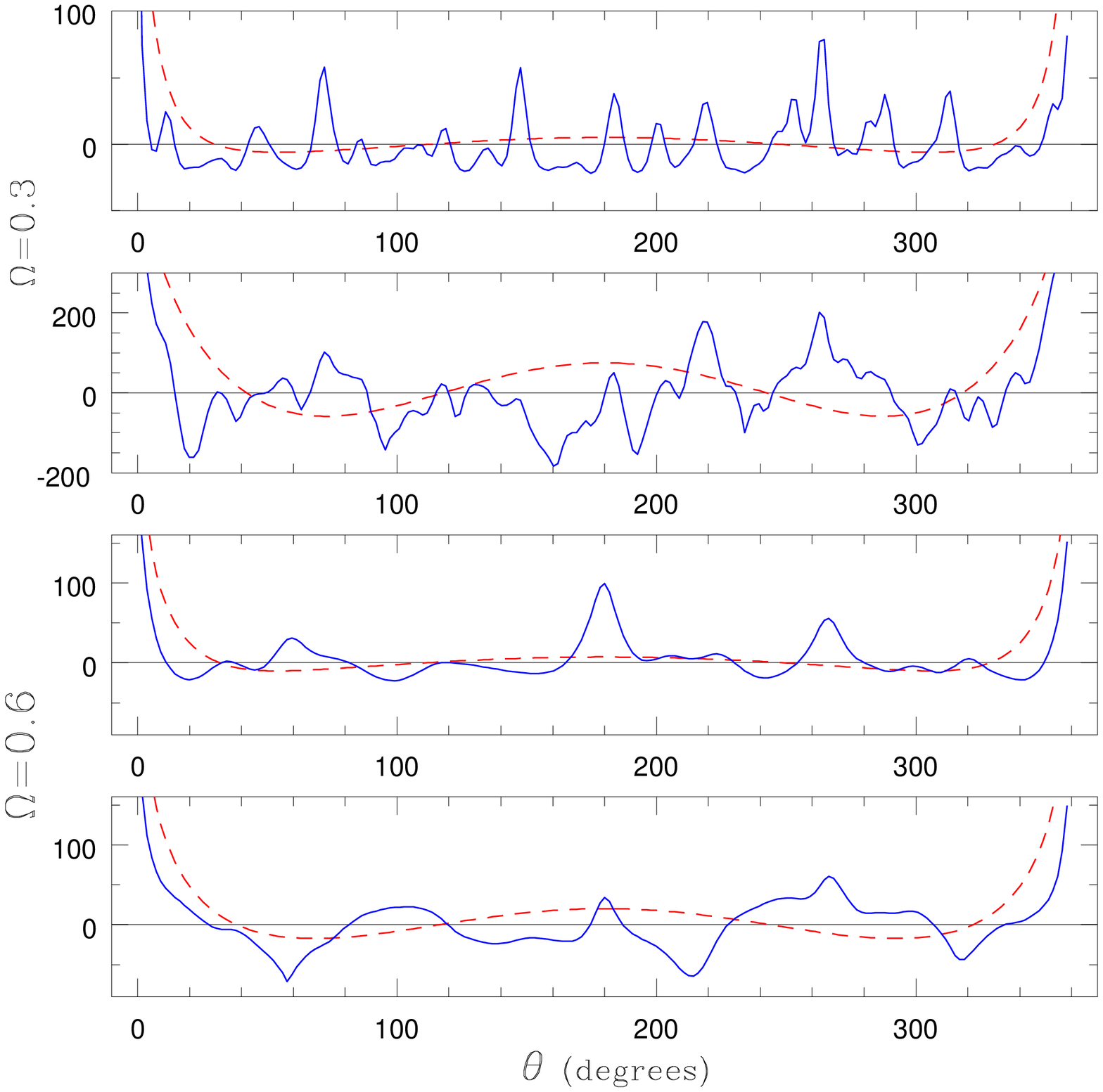}}
\caption{ This shows the behavior of the correlation function of the
CMB temperature on a great circle in the sky in a CH model
(m004(-5,1)). The solid curves in the first and third panels show
$C(\theta)$ for the surface term in the Sachs-Wolfe (NSW) effect for
$\Omega_0=0.3$ \& $0.6$, respectively.  They reflect the spatial
correlation along a circle on the sphere of last scattering (SLS). The
number of peaks in $C(\theta)$ matches the number of Dirichlet domains
that the circle intersects. The dashed curves show corresponding
results for the simply connected infinite $\hm$ models with
$\Omega_0=0.3$ \& $0.6$. The second and fourth panels are analogous to
the first and third panels, where the integrated Sachs-Wolfe (ISW)
effect is included.  In this more physically correct case the sharp
NSW peaks have been diluted by the ISW contributions. However, ISW
induces new features, in particular, note the appearance of strong
negative correlations in panels two and four.}
\label{ctheta_line}
\end{figure}

\begin{figure}[p]
\caption{ This shows two full-sky CMB anisotropy maps, plotted as
pairs of $180^\circ$ diameter hemispherical caps, one centered on the
South Galactic Pole (SGP) and one on the North (NGP). They are one of
an infinite number of possible random realizations based on computed
pixel-pixel correlation matrix for the model in question. Both surface
and integrated Sachs-Wolfe effects have been included. The power was
normalized to best match the {\sc cobe} data.  In contrast to
Figs.~\ref{fig:v_isw} and \ref{fig:m_isw} the maps are not optimally
filtered. The model labels $L(arge)CH$ and $S(mall)CH$ refer to the CH
models v3543(2,3) and m004(-5,1), respectively. (The model number
associated with the topology corresponds to that of the census of CH
spaces from the Geometry center, Univ. of Minnesota; $SCH$ is one of
the smallest and $LCH$ is one of the largest spaces in the
census). The value of $\Omega_0$ in each was chosen so that $V_{\!\cal
M}\sim V_{\sc sls}$. The matched pairs of circles expected if the CMB
anisotropy is dominated by the surface terms~\cite{circles} are
superimposed on the map for each model. Each pair is labeled by the
same number centered on the circles (and the same colour).
The relative phase is shown by
identified points marked by a diamond and a triangle on each circle in
a pair.  For clarity, we show only the eight largest pairs out of $35$
for the $LCH_{\Omega_0=0.6}$ case. Even at {\sc cobe--dmr} resolution,
we find the cross-correlation between the temperature along matched
circles is very good in the $SCH_{\Omega_0=0.9}$ model.  The ISW
contribution is larger at $\Omega_0=0.6$, and the cross-correlation
coefficients are systematically smaller for $LCH_{\Omega_0=0.6}$
circle pairs.  The contours are linearly spaced at $30~\mu {\rm K}$
steps. }

\label{fig:circles}
\end{figure}

\subsection{CMB anisotropy correlation function}
\label{ssec_res_cmbcorr}

In standard cosmological models based on a topologically trivial
space, such as $\hm$, the observed CMB photons have propagated along
radial geodesics from a $2$-sphere of radius $R_{\sc ls}$ (that we
refer to as the sphere of last scattering, SLS) centered on the
observer. The same picture also applies to CH models when the space is
viewed as a tessellation of the universal cover.  If the compact space
fits well within the SLS, the photons propagate through a lattice of
identical domains. As a consequence, strong correlations build up
between CMB temperature fluctuations observed in widely separated
directions.  The correlation function $C (\hat q, \hat q^\prime)$ is
anisotropic and contains characteristic patterns determined by the
photon geodesic structure of the compact manifold. These correlations
persist even in CH models whose Dirchlet domains are comparable to or
slightly bigger than the SLS. This is the key difference from the
standard models where $C (\hat q, \hat q^\prime) $ depends only on the
angle between $\hat q$ and $\hat q^\prime$ and generally falls off
with angular separation.

In Figure\,\ref{ctheta_line}, the complex behavior of the correlation
function is illustrated with the example of the small CH model
(m004(-5,1)). The SLS encompasses $\approx 150$ domains for
$\Omega_0=0.3$ and $\approx 20$ domains for $\Omega_0=0.6$ and is
comparable to the size of one domain for $\Omega_0=0.9$. The angular
correlation along any arbitrary great circle in the sky in lower
$\Omega_0$ models shows distinct peaks as one encounters repeated copies
of the Dirichlet domain. The peaks are more pronounced when one
considers only the surface terms of eq.(\ref{dTSW}) in the CMB
anisotropy. Including the line of sight ISW contribution tends to
smear the peaks but adds its own characteristic
features~\cite{us_inprep}.

Consider compact universe models which are small enough such that the
SLS does not completely fit inside one domain, $R_{\sc ls} > R_<$.
The CMB temperature is expected to be identical along pairs of circles
if temperature fluctuations are dominated by the surface terms at the
SLS~\cite{circles}. We identify these matched circles on the sky in
our models and check the extent of cross correlation seen at the
angular resolution of {\sc cobe--dmr}. \Fref{fig:circles} show the
matched pairs of circles in two (``small'' and ``large'') CH models
superimposed on random realizations of the theoretical sky generated
from the computed pixel-pixel correlation matrix for these models. The
models were chosen to have a volume comparable to the volume $V_{\sc
sls}$ within the SLS. Even with the coarse pixelization of {\sc
cobe--dmr}, we do see fairly good cross-correlation in the CMB
temperature along matched circles in our realizations. Again, the
pattern of correlated circles is more pronounced when the surface
terms are dominant, as happens in the small model ($SCH$) with
$\Omega_0=0.9$.  The cross-correlation coefficient $\rho$ between the
circles is in the range $\rho \approx 0.6-0.95$ in this case, whereas
it is in the range $\rho \approx 0.2-0.6$ for the large model ($LCH$)
with $\Omega_0=0.6$.

The circles of identified pixels on the SLS are not the whole
story~\cite{us_inprep}.  Enhanced NSW cross-correlations, but at a
lower level, also exist between all pairs of points on the SLS which
are projected close to each other on the CH manifold. This can be seen
as secondary maxima in the examples in Figure\,\ref{ctheta_line};
these are absent in standard cosmological models. Some features
persist at a detectable level even when the compact universe
encompasses the SLS, \ie $R_{\sc ls} \lta R_<$ , although circles are
absent in this case (the effect dies out for $R_{\sc ls} \ll R_<$).
When the relative ISW component is significant, the geometrical
patterns based on pointwise identifications on the SLS are supplanted
by more complex features arising from identifications between photon
geodesics, \eg the strong negative correlations evident in
Figure\,\ref{ctheta_line}. Correlation features also arise from
the global inhomogeneity of CH spaces since it implies that the {\it
rms} temperature fluctuations varies with location in the sky in a
pattern set by the topology of the space.

The reader's intuition can be sharpened with examples from simpler
topologies. Flat torus models exhibit forward-backward correlation
symmetry along their axes~\cite{star_angel,us_torus}. Patterns in CMB
temperature maps have also been discussed for other flat models in
~\cite{janna}. An example of a hyperbolic model is the non-compact
horn-like space~\cite{sok_star75} for which ~\cite{janna} argue the
CMB sky will exhibit a flat spot.

\subsection{Bayesian analysis constraints from {\sc cobe--dmr} }
\label{ssec_res_prob}

As discussed above, compact models which are not much larger than the
sphere of last scattering tend to show significant correlated patterns
in the CMB sky. To the extent that these correlations are absent in
the {\sc cobe--dmr} data, the models tend to be at odds with
observations. Figures~\ref{fig:v_isw} and \ref{fig:m_isw} compare
theoretical realizations of the CMB anisotropy in the $LCH$ and $SCH$
models with the {\sc cobe--dmr} data. They should be compared with the
`DATA' map in Fig.~\ref{fig:v_isw}, a Wiener filtered picture of the
CMB data, using a theoretical model which best-fits the data. What one
should be noting is the shapes of the patterns and not the specific
locations of the patterns, since these can change from realization to
realization.  The full Bayesian analysis takes into account all
possible realizations.  The incompatibility of models with small
$V_{\cal M}/V_{\sc sls}$ (SCH-$\Omega_0=0.3,0.6$) is visually obvious:
the best fit amplitudes are high which is reflected in the steeper hot
and cold features. Although, SCH-$\Omega_0=0.9$ and LCH-$\Omega_0=0.6$
do not appear grossly inconsistent, the intrinsic anisotropic
correlation pattern is at odds with the data statistically.
\begin{figure}[p]
\caption{ The figure consists of a column of three CMB sky-maps
showing a pair of $140^\circ$ diameter hemispherical caps each,
centered on the South (SGP) and North (NGP) Galactic Poles,
respectively. The top map labeled DATA, shows the {\sc cobe--dmr}
53+90+31 GHz A+B data after Wiener filtering assuming a standard CDM
model, normalized to {\sc cobe}. The next two maps are one random
realization of the CMB anisotropy in v3543(2,3) -- our choice of a
$L(arge)CH$ model example, for $\Omega_0=0.6~\&~0.8$ based on our
theoretical calculations of $C(\hat q,\hat q^\prime)$ convolved with
the {\sc cobe--dmr} beam. Both surface and integrated (ISW)
Sachs-Wolfe effects have been included in $C(\hat q,\hat
q^\prime)$. No noise was added. The power was normalized to best match
the {\sc cobe} data. The theoretical sky was optimally filtered using
the {\sc cobe} experimental noise to facilitate comparison with data.
LCH with $\Omega_0=0.8$ is compatible with the data with a suitable
choice of orientation while $\Omega_0=0.6$ is ruled out (See
Table~\ref{tab:logprob}). For all the maps in Figs.~\ref{fig:v_isw}
and \ref{fig:m_isw}, the average, dipole and quadrupole of the $\vert b
\vert > 20^\circ$ sky were removed and a $20^\circ$ Galactic latitude
cut was used, with extra cuts to remove known regions of Galactic
emission proposed by the {\sc cobe} team accounting for the ragged
edges. The contours are linearly spaced at $15~\mu {\rm K}$
($\Delta T/T=0.55 \cdot 10^{-5}$) steps.
The maps have been smoothed by a $1.66^\circ$ Gaussian filter.}
\label{fig:v_isw} 
\end{figure}
\begin{figure}[p]
\caption{ The three CMB sky-maps showing a pair of $140^\circ$
diameter hemispherical caps each, centered on the South (SGP) and
North (NGP) Galactic Poles, are analogous to the lower two plots in
Fig.~\ref{fig:v_isw} but for the $S(mall)CH$ model m004(-5,1).  The
fact that $\Omega_0=0.3~\&~0.6$ models above are strongly ruled out by
the {\sc cobe} data is obvious visually, while the $\Omega_0=0.9$ can
be excluded on the basis of our Bayesian analysis.  (See
Table~\ref{tab:logprob}.)}
\label{fig:m_isw} 
\end{figure}

On each of our selection of models, we have carried out a fully
Bayesian analysis of the probability of the model given the {\sc
cobe--dmr} 4yr data. In Table~\ref{tab:logprob}, we present the {\em
relative likelihood} of the selected models to that of the infinite 
$\hm$ model with the same $\Omega_0$.\label{logprob_tab}
\begin{table}[htb]
\caption{The Log-likelihoods of the compact hyperbolic models relative
to the infinite models with same $\Omega_0$ are listed. The
probabilities are calculated by confronting the models with {\sc
cobe--dmr} data. The values quoted are likelihoods marginalized
over the amplitude of the initial power spectrum. The volume within
the sphere of last scattering (SLS) relative to the volume of the
compact models of the universe ($V_{\rm SLS}/V_{\!\cal M}$) is listed.
The three columns of logarithm of likelihood ratios ${\cal L}/{\cal
L}_0$ correspond to the best, next best and the worst values that we
have obtained amongst $24$ different rotations of the compact space
relative to the sky. The number $\nu$ in brackets gives the
conventional, albeit crude, translation of the probabilities to a
Gaussian likelihood ${\cal L}/{\cal L}_0 \sim \exp[-\nu^2/2]$. Only
the last model in a specific orientation appears to be consistent with
the {\sc cobe--dmr} data.}
\begin{indented}
\item[]
\begin{tabular}{@{}cccccc}
\br
CH Topology &$\Omega_0$&$V_{\rm SLS}/V_{\!\cal M}$& 
\centre{3}{Log of Likelihood Ratio (Gaussian approx.)}\\
 & & &\crule{3}\\
$[{\cal V}_{\!\cal M},R_>/d_c]$& & & \centre{3}{Orientation}\\
 &      &  &`best'&`second best'&`worst'\\
\mr
 &0.3& 153.4 & -35.5 (8.4$\sigma$) & -35.7 (8.4$\sigma$)
& -57.9 (10.8$\sigma$)\\
{\bf m004(-5,1)} &0.6&19.3& -22.9 (6.8$\sigma$) & -23.3 (6.8$\sigma$)
& -49.4 ( 9.9$\sigma$)\\
$[0.98,0.75]$  &0.9&1.2 & -4.4  (3.0$\sigma$) & -8.5  (4.1$\sigma$)
& -37.4 ( 8.6$\sigma$)\\
\mr 
{\bf v3543(2,3)} & 0.6&2.9 & -3.6 (2.7$\sigma$) & -5.6 (3.3$\sigma$)
& -31.0 (7.9$\sigma$) \\
$[6.45,1.33]$& 0.8 &0.6&2.5 (2.2$\sigma$) & -0.8 (1.3$\sigma$)
& -12.6 (5.0$\sigma$) \\
\br
\lbl{tab:logprob}
\end{tabular}
\end{indented}
\end{table}
(The {\sc cobe} data alone does not strongly differentiate between the
infinite hyperbolic models with different $\Omega_0$.) The anisotropy
of the theoretical $C(\hat q,\hat q^\prime)$ causes the likelihood of
compact models to vary significantly with the orientation of the space
with respect to the sky, depending on how closely the features in the
single data map available to humans match (or mismatch) the pattern
predicted in $C(\hat q,\hat q^\prime)$. Some optimal orientations may
also have the ``ugly'' correlation features hidden in the Galactic
cut.  We analyzed $24$ different orientations for each of our models
and found that only the model with $V_{\cal M} > V_{\sc sls}$
(LCH-$\Omega_0=0.8$) cannot be excluded. (At one orientation this
model is even preferable to standard CDM; this raises a question of
the statistical significance of any detection of intrinsic anisotropy
of a space when only a single realization of data is available.)

Similar conclusions were reached by some of the authors (JRB, DP and
I.~Sokolov~\cite{us_torus}) for flat toroidal models.  Comparison of
the full angular correlation (computed using the eigenfunction
expansion, eq.(\ref{xi1})) with the {\sc cobe--dmr} data led to a much
stronger limit on the compactness of the universe than limits from
other methods~\cite{star_angel,tor_refs}. The main result of the
analysis was that $V_{\sc sls}/V_{\!\cal M} < 0.4$ at $95\%~CL$ for
the equal-sided $3$-torus. For $3$-tori with only one short dimension
(or for the non-compact $1$-torus), the constraint on the most compact
dimension is not quite as strong because the features can be hidden in
the ``zone of avoidance'' associated with the Galactic cut.

Although our results strongly indicate that manifolds with small
$V_{\!\cal M}/V_{\sc sls}$ are unlikely to survive confrontation with
the {\sc cobe--dmr} data, we emphasize that when this ratio becomes of
order unity very interesting correlation patterns appear which, for
certain orientations, may even be preferred by the {\sc cobe--dmr}
data. Since manifolds with large $V_{\!\cal M}/d_c^3$ allow low values
of $\Omega_0$ while still avoiding small $V_{\!\cal M}/V_{\sc sls}$,
even large mean curvatures may be compatible, and certainly the
$\Omega_0 \sim 0.3-0.6$ values that some of the astrophysical data
currently prefers can still be nicely accommodated within the CH
framework. \footnote{ CH models with non-zero cosmological parameter,
$\Lambda$, and associated density, $\Omega_\Lambda$ are certainly
possible. The {\sc cobe--dmr} data by itself does not strongly
constrain $\Omega_\Lambda$, \eg~\cite{bondLH}, and the constraints
presented here would roughly correspond if $\Omega_0$ is replaced by
$\Omega_0 + \Omega_\Lambda$, although there will be be quantitative
differences.} Of course, there are many more manifolds for which
$R_{<}$ may be small but $R_{>}$ large, for whom the constraints will
be even more dependent upon how specific orientations may deliver
unwanted correlation features into the zone of avoidance, as the
asymmetric 3-torus results reveal, or line up predicted correlation
patterns with chance upward or downward fluctuations in the data. When
$R_{<}$ is large compared to $R_{\sc ls}$, we expect the results will
quickly converge towards the usual infinite hyperbolic manifold
results. The intermediate terrain still encompasses ample scope for
interesting topological signatures to be discovered within the
CMB. Although our methods are quite general, testing all manifolds in
the SnapPea census this way is rather daunting, and there are
countably infinite manifolds not yet prescribed. What may be more
promising for discovery are specialized statistical indicators, which
are less sensitive than the full Bayesian approach we have used here,
but not as manifold sensitive; {\it e.g.}, the statistical techniques
which exploit the high degree of correlations along circle pairs that
\cite{circles} have emphasized, and Fig.~\ref{fig:circles}
reveals. Maps like we have constructed will be necessary to test the
statistical significance of such methods. We also note that
dramatically increasing the resolution beyond that of {\sc cobe--dmr}
is quite feasible with current computing power using our techniques.

\section{Acknowledgements}
We thank Jeff Cheeger for illuminating discussions.  We also thank the
Geometry center of the University of Minnesota for making a lot of
useful material available on their website, in particular, the SnapPea
package. We thank the anonymous referee for pointing out related
literature~\cite{minvol,aur_stein92}.

\def\prd{{Phys.~Rev.~D}}
\def\prl{{Phys.~Rev.~Lett.}}
\def\apj{{Ap.~J.}}
\def\apjl{{Ap.~J.~Lett.}}
\def\apjsuppl{{Ap.~J.~Supp.}}
\def\mnras{{M.N.R.A.S.}}

\end{document}